\documentclass{aastex62}
\usepackage{amsmath}
\usepackage{bm}
\graphicspath{{./}{figures/}}

\received{February 12, 2018}
\revised{N/A}
\accepted{N/A}
\submitjournal{ApJ}

\shorttitle{On Rotation Curve Analysis}
\shortauthors{Kadowaki}
\begin{document}

\title{On Rotation Curve Analysis}

\correspondingauthor{Kevin Kadowaki}
\email{kadowakk@uci.edu}

\author{Kevin Kadowaki}
\affil{University of California, Irvine}

\begin{abstract}

An analysis of analytical methods for computing galactic masses on the basis of observed rotation curves \citep{saari15} is shown to be flawed.

\end{abstract}

\keywords{celestial mechanics --- 
galaxies: kinematics and dynamics --- methods: analytical}


\section{Introduction}

There are two standard responses to the discrepancy between observed galactic 
rotation curves and the theoretical curves calculated on the basis of luminous 
matter:  postulate dark matter, or modify gravity.  Most physicists accept the 
former as part of the concordance model of cosmology; the latter encompasses a 
family of proposals, of which MOND is perhaps the best-known example.  Don Saari, 
however, claims to have found a third alternative: to explain this discrepancy as 
a result of approximation methods which are unfaithful to the underlying
Newtonian dynamics.  If he is correct, eliminating the problematic approximations 
should allow physicists and astronomers to preserve the validity of Newtonian dynamics in 
galactic systems without invoking dark matter.

As physicists and astronomers have found a wide range of other empirical tests for the existence 
of dark matter (e.g., \citep{zwicky37,clowe06,spergel07}), Saari's criticism does 
not (by itself) give us reason to doubt the existence of dark matter.  
Nevertheless, there are several good reasons to address Saari's argument.  In the 
context of physics, rotation curves remain a key source of evidence for dark 
matter at the galactic scale---other evidence, such as CMB anisotropies, operates 
at the cosmological scale.  Philosophically, this example brings to bear a number 
of issues surrounding the use of approximation and idealization in physical 
theories.  In both cases, a successful skepticism regarding the connection 
between galactic rotation curves and dark matter would require us to reevaluate 
these methods and our expectations for the future of cosmological research.  I 
will endeavor to show that such a radical reevaluation is unnecessary---at least 
on the basis of the considerations presented in \citep{saari15}.

The paper will be divided into three main sections.  In the first section, I will 
outline Saari's argument.  In the second section, I will show that Saari's 
explanation for the effectiveness of his counterexample is incorrect.  In the 
final section, I will show that his counterexample fails to address the standard 
methods used to model galaxies.

Some of the results in this paper have been derived using the HEALPix
\citep{gorski05} package.

\section{Saari's Argument}

As a typical galaxy is made up of $\approx 10^{11}$ stars, treating it as a 
$10^{11}$-body problem would be computationally prohibitive.  As a result, 
physicists analyze rotation curves by assuming that the galaxy can be 
approximated by a continuum distribution.  Saari notes that in symmetric 
settings, the mass $M(r)$ contained up to radius $R$ is given by
\begin{equation} M(r)=\frac{rV(r)^2}{G}\end{equation}
where $V(r)$ is the rotational velocity of a star at distance $r$.  When we apply 
this to observed rotation curves, we note that the flattening of the rotation 
curve implies that at the edge of galaxies $M(r)\propto r$.

Saari contends that this continuous approximation does not accurately 
track the distribution of matter in at least some large $N$-body systems, and he 
proves this by constructing a family of large $N$-body systems for which the 
approximation is demonstrably incorrect.  In this section, I will give a sketch 
of the proof.  The full version can be found in \citep{saari15}.

\begin{figure}
  \centering
  \includegraphics[width=40 mm]{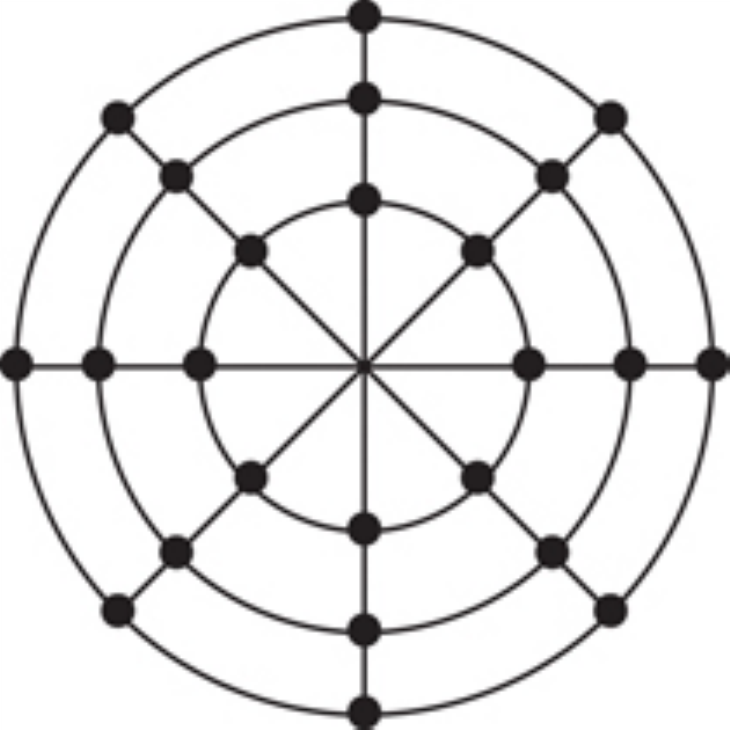}
  \caption{From \cite{saari15}.  A spiderweb configuration for $n=3$, $k=4$.}
  \label{fig:spiderweb}
\end{figure}

We define a {\em spiderweb configuration} to be a system of $N=2nk$ point masses 
in a plane, positioned such that they resemble a spiderweb---see Figure $1$ for 
an example of $n=3$, $k=4$.  More precisely, a series of $n$ concentric rings are 
intersected by $k$ evenly-spaced spokes, and a body is placed at each 
intersection of a ring and a spoke.  The mass of each body on a given ring is the 
equal to all other masses on that ring; generally, however, masses from separate 
rings will differ.  Conventionally, we will let $r_i$ refer to the radius of the 
$i$th ring, and we will let $m_i$ refer to the mass of a single body on the $i$th 
ring.  Note that the symmetry of the configuration guarantees that the net force 
on each body will be in the radial (or anti-radial) direction and that every mass 
in a given ring will experience a force of the same magnitude.  As a result, the 
position and acceleration vectors satisfy
\begin{equation} \bm{r}''_i=\lambda_i \bm{r}_i\end{equation}
for some scalars $\lambda_i$.

Saari's proof exploits the properties of a set of special spiderweb 
configurations, which I will call {\em Saari configurations}.  A Saari 
configuration is defined as a spiderweb configuration with the additional 
property that $\lambda_i=\lambda_j$ for all $i,j$.  Thus, in a Saari 
configuration, for some constant $\lambda$,
\begin{equation}\bm{r}''_i=\lambda \bm{r}_i\end{equation}

Physically, a Saari configuation is one in which, given the proper initial 
velocities to the various bodies, the system will rotate as a rigid body; i.e., 
all bodies will undergo uniform circular motion about the origin with the same 
angular velocity $\sqrt{\lambda}$.  This will obviously not be true of generic 
spiderweb configurations---though the symmetry of the spiderweb configuration 
always guarantees that the acceleration of each body will be purely radial, 
generally $\lambda_i\not=\lambda_j$ for $j\not=i$.

Saari first proves, given $n$ rings and $2k$ masses per ring, that for any 
specification of masses $m_i$ there exists values of $r_i$ such that the 
resulting spiderweb configuration is a Saari configuration.  Furthermore, these 
values $r_i$ are unique up to a scale factor.  The full details of this proof are 
not essential, but a motivating sketch will suffice for our purposes.  Consider 
the case of $n=2$ rings.  Fix $r_1$, and define $x_2=r_2-r_1$.  We can then 
define $\lambda_1(x_2)$ and $\lambda_2(x_2)$---these functions will be monotonic, 
as force that each ring exerts on the other will strictly decrease as the 
distance between them increases.  We note that by making $x_2$ arbitrarily small, 
the masses on the outer ring exert arbitrarily large forces on the masses in the 
inner ring---thus, as $x_2\rightarrow 0$, $\lambda_1\rightarrow \infty$ and 
$\lambda_2\rightarrow -\infty$.  As $x_2$ is made arbitrarily large and the 
forces between the inner and outer rings approach $0$, the inter-ring forces on 
the inner and outer rings approach a negative constant and zero, respectively---
thus, as $x_2\rightarrow \infty$, $\lambda_1\rightarrow \lambda'$ and $\lambda_2
\rightarrow 0$ for some $\lambda'<0$.  The continuity and limits of these curves 
require that there exists some $x_2'$ such that $\lambda_1(x_2')=
\lambda_2(x_2')$; the fact that these curves are monotonic implies that this 
value is unique.  The proof for $n>2$ is more complicated, but the general 
strategy is similar.

With this proof in hand, Saari presents his counterexample.  Let $n$, $k$, be 
suitably large, and for all $i$ let $m_i=1/2k$, so that the total mass of each 
ring is normalized to $1$.  By the previous proof, there are values of $r_i$ such 
that this configuration is a Saari configuration; scale these values such that 
the spacing between each ring is at least unit (i.e., $r_{i+1}-r_i>1$ for all 
$i<n$).  This spacing ensures that for all $i$, $r_i\le i$.

Saari configurations all rotate with uniform angular velocity---that is, 
$V(r)=\sqrt{\lambda} r$.  Thus, the Equation 1 approximation predicts that
\begin{equation} M_c(r)=\frac{\lambda r^3}{G}\end{equation}
That is, $M_c(r)\propto r^3$, where $M_c(r)$ denotes the estimate derived from 
the approximation.  But by construction, $r_i\ge i$, and because the mass of each 
ring is equal to 1, $M(r_i)=i$.  Thus, for the constructed Saari configuration, 
the actual mass distribution must be consistent with the relation $M(r_i)\le 
r_i$; at most, $M(r)\propto r$.

This difference between the actual mass distribution and the prediction made by 
Equation 1 is not trivial.  As Saari points out, it is impossible to make a cubic 
equation approximate a linear equation.  Moreover, we can construct these 
configurations for arbitrarily large $n$, so even if we were able to convince 
ourselves that the cubic prediction approximated the linear distribution in some 
limited domain, we could not extend this to arbitrarily larger domains.

Obviously, this family of configurations does not resemble a galaxy---unlike the 
observed rotation curves of galaxies, the rotation curves of these configurations 
increase linearly.  As such, one might suppose that this whole line of argument is 
of questionable relevance until it can be applied to mass profiles that more closely 
resemble galaxies.  This objection is feasible, but unnecessary.  We can grant that 
Saari has successfully cast doubt on the ability of Equation 1 to accurately model 
large $N$-body simulations in general.  Saari himself acknowledges that his 
counterexample does not definitively show that Equation 1 is a bad approximation 
in the case of galaxies, but he argues that some additional justification must 
be given for its use in the light of his counterexamples.  The particular 
nature of these justifications are related to his interpretation of 
his proof, which I will discuss in the next section.

\section{Analysis of Saari's Proof}
In this section, I will {\em not} be arguing that Saari has made any errors in 
the construction of his proof.  Insofar as the technical details of his proof are 
concerned, he is entirely correct.  I will rather be arguing that the {\em 
reasons} Saari gives for the effectiveness of his counterexample are incorrect.

Broadly speaking, Saari attributes the failure of Equation 1 to a 
``reductionist'' approach:  by analogy to Arrow's theorem, he argues that by 
simplifying a problem into individual component parts and solving each part 
separately, information that is otherwise preserved by a holistic treatment of 
the system is lost.  In the specific case of dark matter, he argues that, by 
treating galaxies as continuous distributions instead of discrete $N$-body 
problems, astronomers have smoothed over and ignored the ``connecting links'' 
between bodies \citep{saari16}.  His contention is that when two bodies closely 
approach each other while traveling in their circular orbits, as in Figure 2, the 
inner body (A) `tugs' on the outer body (B) in such a way that the outer body's 
rotational velocity increases, and that this `tugging' is ignored by the 
continuous treatment of the N-body problem \citep{saari15}.

\begin{figure}[h]
  \centering
  \includegraphics[width=55 mm]{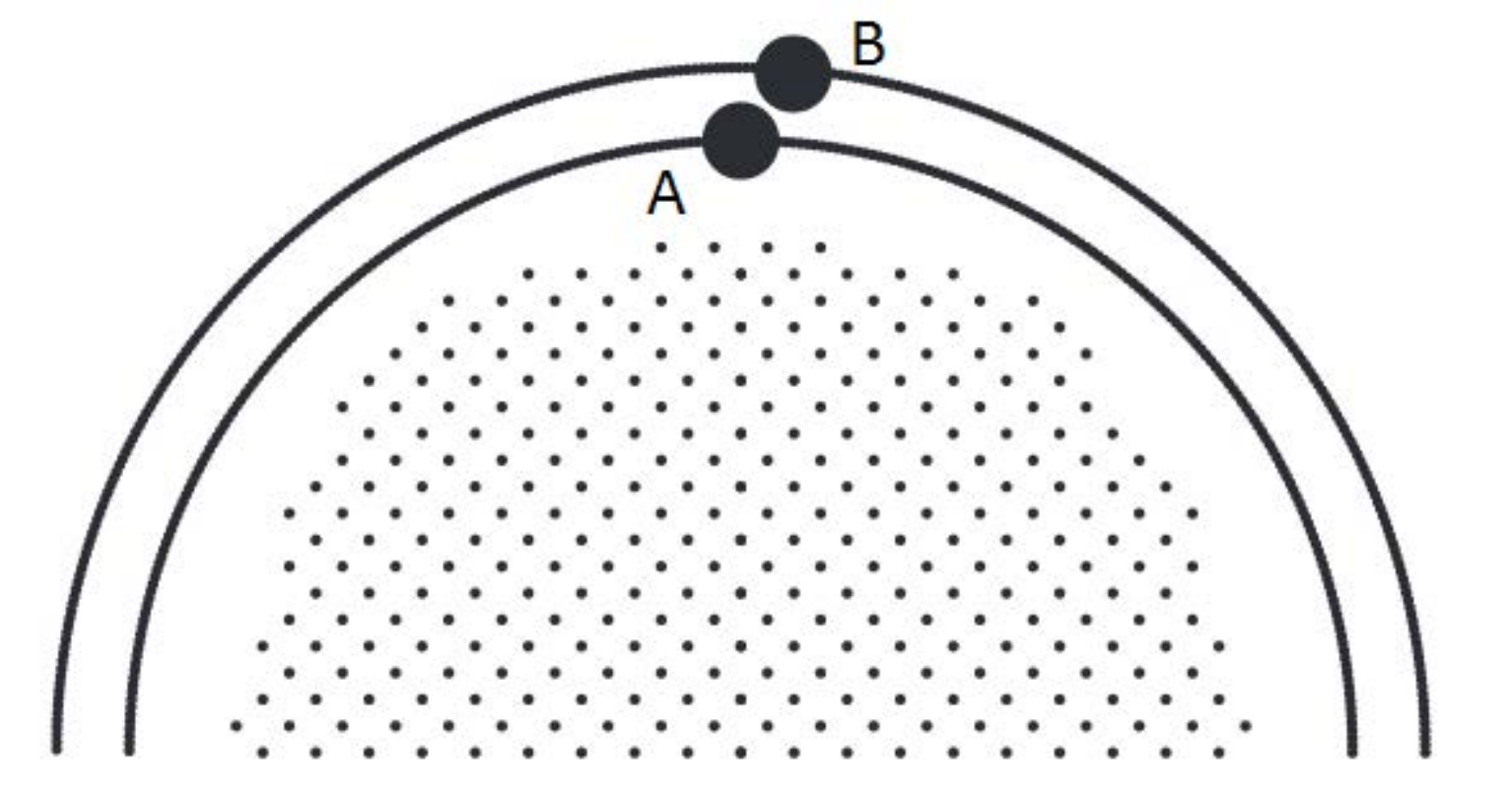}
  \caption{Adapted from \cite{saari15}.}
  \label{fig:tug1}
\end{figure}

As an aside, Saari's invocation of reductionism is of questionable relevance here.  While 
treating a galaxy as a continuum distribution is certainly a simplification of the actual 
system, this is arguably the most `holistic' treatment of the problem.  Indeed, the 
reductionist method of breaking a problem down into its component parts is a more apt 
description of a discrete $N$-body problem, where the principle of superposition allows us to 
sum together the contributions of pairwise component parts without considering more complex 
holistic interactions.  I will focus on the distinction between the analysis of galaxies as 
continuous distributions and as as discrete $N$-body problems.

Equation 1 does not fail to accurately model Saari configurations because it treats a 
discrete system as continuous; it fails because it requires an assumption of spherical 
symmetry and Saari configurations lack this symmetry.  Equation 1 is a consequence of Gauss's 
law, which states that for a closed surface $S$, the flux of the gravitational field $\bm{g}$ 
through $S$ satisfies
\begin{equation} \oint_S \bm{g}\cdot dA = -4\pi GM_{enc} \end{equation}
where $M_{enc}$ is the mass enclosed by the surface $S$.  If we make the additional 
assumption that our mass distribution is spherically symmetric and choose a concentric sphere 
of radius $r$ as our Gaussian surface, then the spherical symmetry implies that $|\bm{g}|$ is 
a constant over the Gaussian surface.  Thus $|\bm{g}|4\pi r^2=4\pi GM(r)$.  If a body is in 
uniform circular motion at radius $r$, its centripetal acceleration is given by $V(r)^2/r$, 
and this simplifies to
\[M(r)=\frac{V(r)^2 r}{G}\]
which is just Equation 1.   Note that the assumption of spherical symmetry was crucial for 
this derivation; without it, we could not simplify the surface integral as we did.

This analysis can be substantiated by two straightforward examples.  If Saari's contention is 
correct, then two claims should follow:  (1) continuous, non-spherically symmetric 
distributions should be well-approximated by Equation 1 and (2) discrete, spherically 
symmetric distributions should {\em not} be well-approximated by Equation 1.  Counterexamples 
to the first claim are not difficult to construct---consider any spherically symmetric 
potential $\phi(x,y,z)$ and transform coordinates $\phi(x,y,z)\mapsto\phi(kx,y/k,z)$ to 
stretch and compress this potential across orthogonal axes.  The resulting potential will 
trace a distribution in which the rotation curves along these axes are significantly 
different and which cannot all be consistent with Equation 1.  With respect to the second 
claim, we will give an example of a discrete, approximately spherically symmetric 
distribution for which Equation 1 is a good approximation.  In the following example, we use 
units where $G=1$.  %Insert a figure illustrating the first counterexample?

Using the HEALPix package\footnote{HEALPix---Hierarchical Equal Area isoLatitude Pixelization 
scheme.  Website:  http://healpix.sourceforge.net} at resolution 4, we discretized a sphere 
into 3072 point masses, each of unit mass.  We nested 20 of these discretized spheres 
concentrically, setting the radius of each sphere $r_i=i$.  Assuming that a generic point 
mass from each sphere was instantaneously undergoing uniform circular motion, we calculated 
the force on each particle and used Equation 1 to estimate $M(r)$.  The estimated and actual 
values of $M(r)$ are plotted in Figure 3; note that, unlike the previous example, Equation 1 
successfully approximates the mass distribution.  Thus, Equation 1 can give good 
approximations of discrete systems if those systems are approximately spherically symmetric.

\begin{figure}[!h]
  \centering
  \includegraphics[width=75 mm]{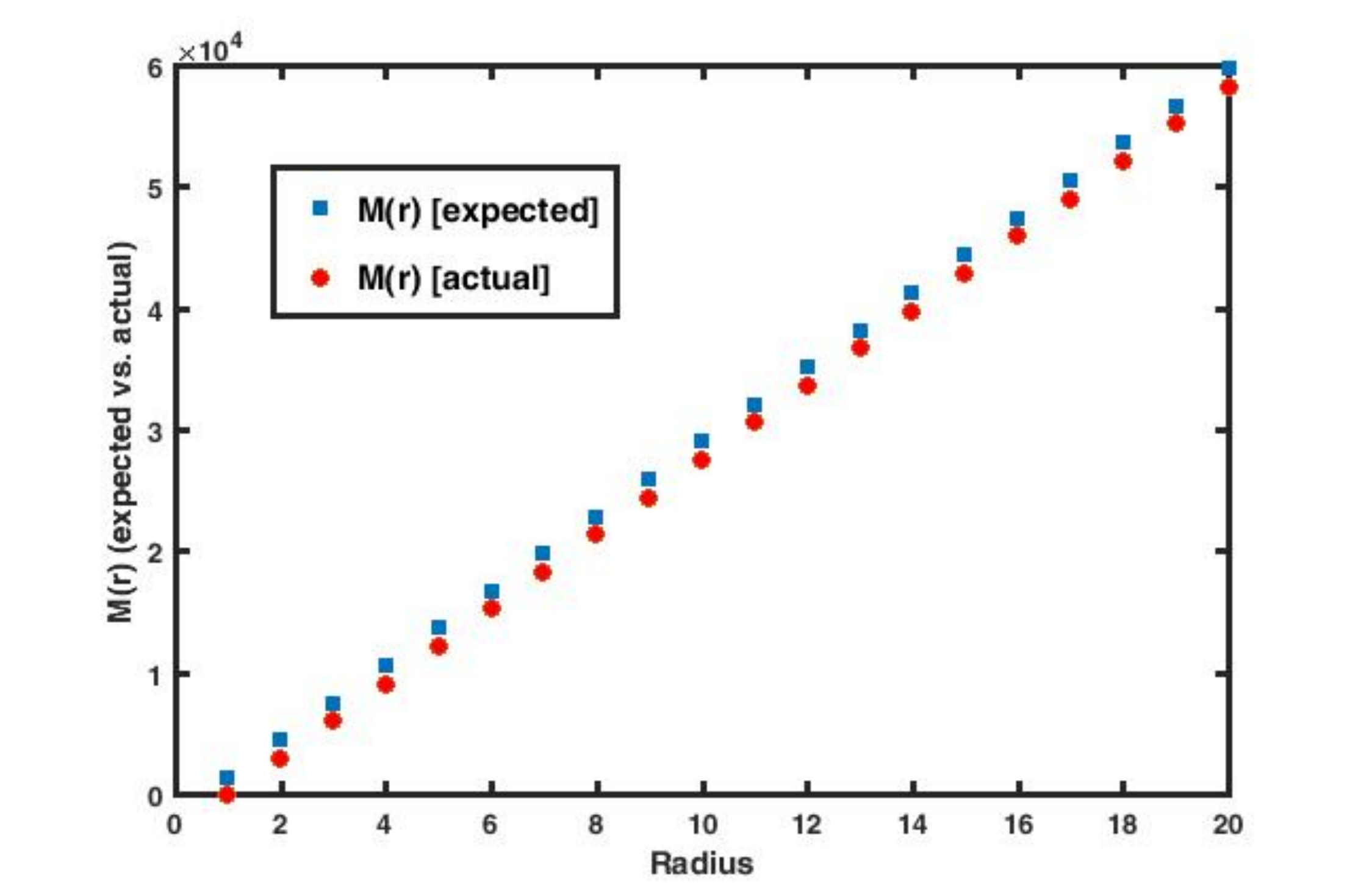}
  \caption{Comparison between expected value for $M(r)$ based on Equation 1 (squares), and the actual value of $M(r)$ (circles).}
  \label{fig:ex3}
\end{figure}

Given that Saari configurations clearly lack spherical symmetry, it is unsurprising that 
Equation 1 fails to approximate Saari configurations.  Of course, despite this 
misinterpretation, Saari's proof still shows that Equation 1 will fail to accurately model 
some distributions that lack approximate spherical symmetry.  As galaxies are not spherically 
symmetric, a full response to his challenge requires us to more closely examine 
the standard methods used to model galaxies.  In the next section, I will argue that 
Saari has misrepresented these methods.

\section{Discussion and Evaluation of Saari's Argument}
In the light of his counterexample, Saari calls for new justifications for the use of Equation 1 and other continuous approximations \citep{saari16}.  But there are a number of misconceptions embedded in this challenge---first and foremost, the notion that Equation 1 is typically used to model galaxies.  Physicists and astronomers are, in fact, aware that galaxies are not spherical, and adjust their methods appropriately.  For example, the landmark paper \citep{rubin70} uses a disk potential to model the galaxy M31; the mass function associated with this model is not Equation 1 but
\begin{equation}
M(r)=\frac{2}{G\pi}\int_0^r\frac{V^2(a)a}{(r^2-a^2)^{1/2}}\ da
\end{equation}
Further examples are not difficult to find, and a broad overview of these methods will make it apparent why.

The procedure is simple.  Assuming non-gravitational pressure terms are negligible, the acceleration of a mass element is purely gravitational.  Assuming stability and rotational symmetry, this acceleration is purely centripetal.  The resulting equation, in terms of the rotational velocity curve, is
\begin{equation}
\frac{V(r)^2}{r}=\nabla\phi
\end{equation}
where $\phi$ is the gravitational potential.  Note that Equation 7 does not have a unique solution, even up to an integration constant; the curve $V(r)$ only constrains the potential gradient in the plane of $z=0$.  As a result, the rotation curve does not give us enough information to calculate a unique mass distribution, and one must make additional assumptions about the form of the potential.  The simplest involve spherical symmetry, but even these range from the familiar point-mass potential to power-law density models.  More complex shapes, such as spheroids \citep{burbidge59} and thin disks \citep{toomre63}, are more common, as these better approximate the shape of galaxies.  And because $\nabla$ is a linear operator, linear combinations of these potential forms will also be solutions; one can model a galaxy as a thin disk with a spheroid bulge in the center \citep{shu71}.  One can even use inhomogeneous spheroids to analyze asymmetric aspects of galaxies, such as the spiral arms.
% Add "e.g., Shu71" instead?

Thus, contrary to Saari's contention, physicists have the tools necessary to account for his alleged counterexample systems.  To impress this point, we will give an example of a spiderweb configuration approximated using a disk potential given by
\begin{equation}
\Phi(R,0)=-4G\int_0^R\frac{da}{\sqrt{R^2-a^2}}\int_a^\infty dR'\frac{R'\Sigma(R)}{\sqrt{R'^2-a^2}}
\end{equation}
where $\Sigma(R)$ is the surface density of the mass distribution.  We considered a configuration with $n=50$ rings and $k=100$ spokes, with $r_i=i$ and $m_i=1$ for all rings. 
After submitting this discrete distribution to a coarse-graining procedure to find the corresponding approximation $\Sigma(R)$, we numerically solved for the potential in the plane using Equation 8.  Figure 4 compares this approximate potential to the actual potential along a generic spoke; note that as the physically salient information is captured by the {\em gradient} of the potential, the continuous estimate is a good approximation despite the extra constant of integration.

\begin{figure}[ht!]
  \centering
  \includegraphics[width=75 mm]{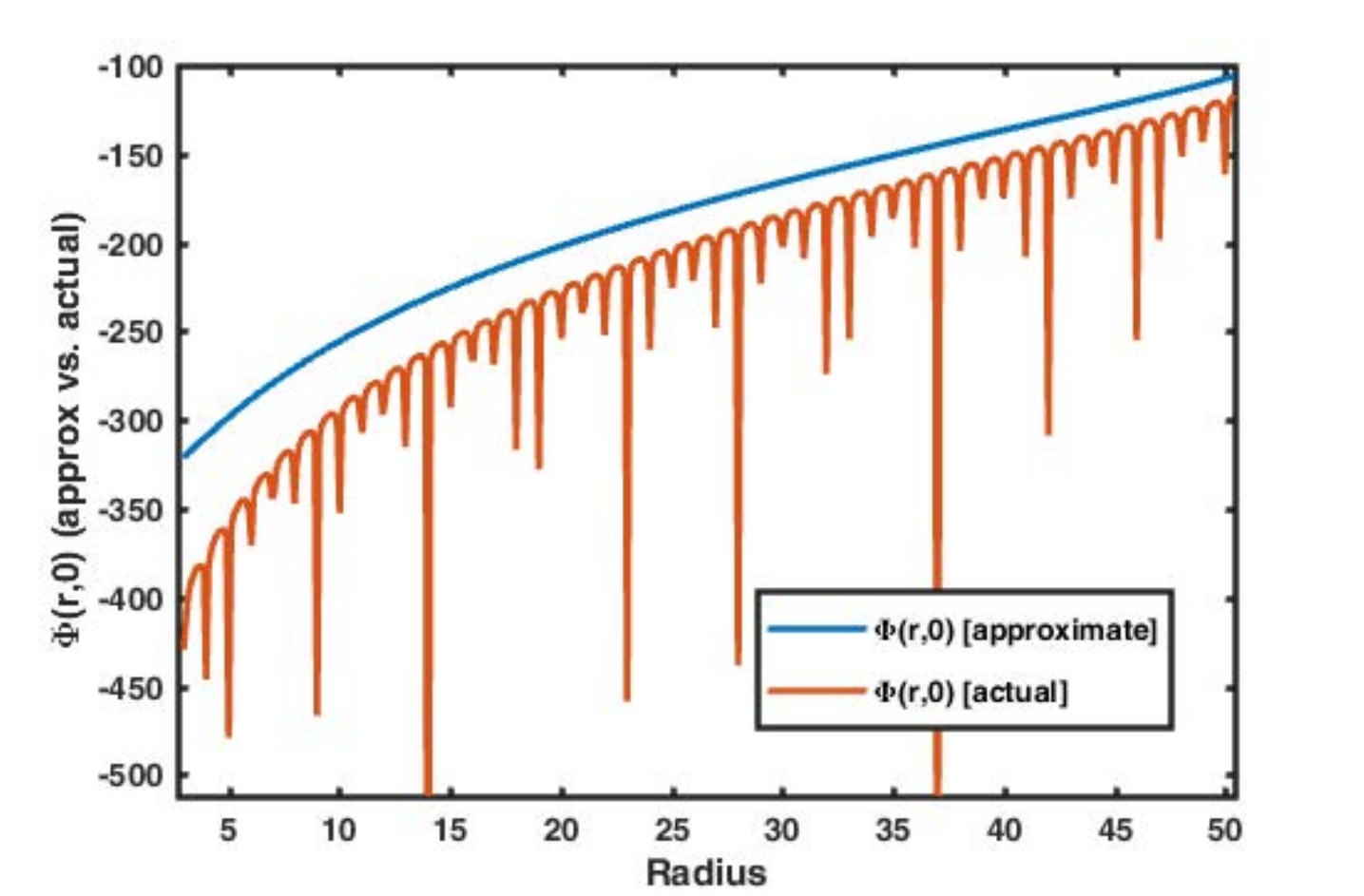}
  \caption{Comparison between the continuous approximation and the actual potential for a 50-ring spiderweb configuration.}
  \label{diskpot}
\end{figure}

Thus, Equation 1 does not play the central role that Saari contends.  His proof does not pose a problem for the standard methods of rotation curve analysis, because it merely addresses an oversimplified straw-man of these methods.  One might interpret Saari's claim more broadly---as a call to justify the general treatment of discrete systems with continuous approximations.  But even this, however, is problematic.  Construed in this broad fashion, this suggests that astronomers and physicists have {\em not} given justifications their use of continuous approximations in the context of galaxy modeling, and this is simply false.

When one assumes that a galaxy can be modeled by a continuous approximation, this continuous approximation smooths over local variations in the gravitational field.  These local variations are especially evident when we consider a collection of point masses such as one of the configurations described above, for the gravitational attraction about these singularities is arbitrarily large.  Saari contends that these local variations are responsible for ``tugging'' effects which will distort the system in ways that the smoothed-over approximation does not account for.  But one can estimate these distortions and, from this estimate, assess the faithfulness of the approximation.  A detailed account can be found in \citep{binney11}, \S 1.2, but we will sketch the process.

One can estimate the amount $\delta v$ by which a star will be deflected in passing by another star as a function of the distance of closest approach $b$; this $\delta v$ represents the the deviation unaccounted for by the continuous approximation.  Based on the density of stars, one can then integrate over $b$ to estimate the total deviation that a typical star will undergo in a single crossing of the galaxy.  Finally, one can estimate the number of times that a star will have to cross the galaxy for this distortion to be severe---that is, for the distortion to be on the order of the original velocity.  The amount of time that it takes for a star to undergo this number of crossings is the relaxation time, and on time scales lower than the relaxation time the system can safely be treated as collisionless and well-approximated by a smooth continuous potential.

Saari may be able to lodge an objection against this estimation, but this would require a targeted analysis of the mathematics involved.  Moreover, this is not the only possible justification for the various modeling decisions that one can make---for example, an entire subbranch of astrophysics is dedicated to analyzing this problem using large N-body simulations.  Of course, these simulations make computation feasible by means of other compromises.  One might, for instance, `soften' the gravitational force at short distances (i.e., let $F(r)\rightarrow 0$ as $r\rightarrow 0$) to avoid the numerical problems caused by the usual divergence of $F(r)$ as $r\rightarrow 0$, even though this introduces some distortions into the simulation.  But a comprehensive objection to the standard methods for galaxy modeling would require a careful treatment of {\em all} these different justifications---and an explanation of how this wide variety of methods have all managed to come to the same allegedly flawed conclusion.

\acknowledgments

I would like to thank Jim Weatherall and Manoj Kaplinghat for their comments and feedback.

\end{document}